# Media-TCP: A Quality-Centric TCP-Friendly Congestion Control for Multimedia Transmission

Hsien-Po Shiang and Mihaela van der Schaar
Department of Electrical Engineering, UCLA, Los Angeles, USA
{hpshiang, mihaela}@ee.ucla.edu

*Abstract—* **In this paper, we propose a quality-centric congestion control for multimedia streaming over IP networks, which we refer to as media-TCP. Unlike existing congestion control schemes that adapt a user's sending rate merely to the network condition, our solution adapts the sending rate to both the network condition and the application characteristics by explicitly considering the distortion impacts, delay deadlines, and interdependencies of different video packet classes. Hence, our media-aware solution is able to provide differential services for transmitting various packet classes and thereby, further improves the multimedia streaming quality. We model this problem using a Finite-Horizon Markov Decision Process (FHMDP) and determine the optimal congestion control policy that maximizes the long-term multimedia quality, while adhering to the horizon-$K$ TCP-friendliness constraint, which ensures long-term fairness with existing TCP applications. We show that the FHMDP problem can be decomposed into multiple optimal stopping problems, which admit a low-complexity threshold-based solution. Moreover, unlike existing congestion control approaches, which focus on maintaining throughput-based fairness among users, the proposed media-TCP aims to achieve quality-based fairness among multimedia users. We also derive sufficient conditions for multiple multimedia users to achieve quality-based fairness using media-TCP congestion control. Our simulation results show that the proposed media-TCP achieves more than 3dB improvement in terms of PSNR over the conventional TCP congestion control approaches, with the largest improvements observed for real-time streaming applications requiring stringent playback delays.**

## I. INTRODUCTION

Transmission Control Protocol (TCP) is the most widely used protocol for data transmission at the transport layer. However, existing TCP congestion control provides dramatically varying throughput that is unsuitable for delay-sensitive, bandwidth-intense, and loss-tolerant multimedia applications [3] (e.g. real-time video streaming, videoconferencing etc.). This is due to the fact that current TCP congestion control aggressively increases the congestion window until congestion occurs, and then adopts an exponential backoff mechanism to mitigate the congestion. The fluctuating throughput results in long end-to-end delays which can easily violate the hard delay deadlines required by various multimedia applications. Hence, numerous multimedia transmission solutions over IP networks adopt User Datagram Protocol (UDP) at the transport layer [27]. However, UDP provides connectionless, unreliable services without guaranteed delivery, which limits the quality of service (QoS) support for multimedia applications at the transport layer. Therefore, multimedia applications need to rely on error resilience [8], forward error correction [9][10] and/or source coding rate control solutions [11][27], which need to be implemented at the application layer to achieve a desirable streaming quality. Moreover, the lack of congestion control mechanisms in UDP can lead to severe network congestion. Therefore, a significant body of existing multimedia streaming research over the past decade has focused on applying UDP-based congestion control that are TCP-friendly [4], which are being standardized as the Datagram Congestion Control Protocol (DCCP) [28]. However, these solutions often ignore the specific characteristics and requirements of multimedia applications, thereby leading to a sub-optimal performance for these applications.

Importantly, multimedia applications have several unique characteristics which need to be taken into account when designing a suitable congestion control mechanism. First, multimedia applications are loss-tolerant, and graceful quality degradation can be achieved if multimedia packets having lower distortion impacts are not received. Hence, various scheduling strategies [18][19] were proposed to optimize the received multimedia quality for multimedia streaming by prioritizing packets for transmission over error-prone IP networks. Such solutions, which explicitly consider multimedia packets' distortion impacts, have also been adopted in order to improve the performance of congestion control mechanisms for multimedia applications [7]. Secondly, multimedia applications are delay-sensitive, i.e. multimedia packets have hard delay deadlines by which they must be decoded. If multimedia packets cannot be received at the destination before their delay deadlines, they should be purged from the senders' transmission buffers to avoid wasting precious bandwidth resources. Third, in order to remove the temporal correlation existing in the source data, multimedia data are often encoded interdependently using prediction-based coding solutions (as in [20][21]). This introduces sophisticated dependencies between multimedia packets across time. Hence, if a multimedia packet is not received at the destination before its delay deadline, all the packets that depend on that packet should be purged from the transmission buffer to avoid unnecessary congestion, since these packets are not usable at the decoder side.

### A. Limitations of current trasnport layer solutions for multimedia transmission

Supporting real-time multimedia transmission over IP networks is an important, yet challenging problem. Various approaches have been proposed to adapt the existing transport layer protocols such that they can better support delay-sensitive, loss-tolerant multimedia applications. However, most current approaches still exhibit several key limitations.

**1) Multimedia quality unaware adaptation**. Conventional transport layer congestion control approaches are application-agnostic, meaning that they merely attempt to avoid the network congestion by adjusting the sending rates, without considering the impact on the application's performance. For example, many TCP-friendly approaches apply analytical models [2] on the long-term TCP throughput and adapt the sending rate to the periodically updated TCP throughput [3][4]. These model-based approaches aim to optimize the bandwidth utilization, which may fail to maximize the multimedia application performance (e.g. video quality) since they do not consider multimedia characteristics, such as distortion impact, delay deadline, etc.

**2) Flow-based models for multimedia traffic without delay consideration**. Various approaches are proposed to adapt multimedia applications to the available TCP throughput by applying rate-distortion optimization [11], forward error coding (FEC) [9][10], or frame dropping [25]. These solutions often adopt flow-based models for multimedia traffic that only consider the high-level flow rate (e.g. the average rate and peak rate of the flow/frame [27]). They do not explicitly consider the specific distortion impact and delay deadline of each packet, as well as the interdependencies existing among packets of multimedia applications. Hence, these congestion control approaches only provide suboptimal solutions for multimedia transmission [18][29].

**3) Myopic adaptation.** Prediction/estimation of the network condition is widely used in congestion control mechanisms based on network information feedback, e.g. in [4][11]. However, these solutions adapt the congestion window myopically, i.e. based only on the current network condition. Considering the packets' delay deadlines and dependencies in the transmission buffer, the congestion window size not only impacts the immediate multimedia quality, but also impacts the available packets in the buffer for future transmission. Hence, it is important to consider not only the instantaneous multimedia quality, but also how the immediate congestion window size impacts the long-term expected quality in the subsequent time slots. In [12], it was shown that the quality can be improved by allowing temporary violation of the TCP-friendliness, while later compensating the congestion control to maintain *long-term* TCP-friendliness. They proposed a joint source rate control and QoS-aware congestion control scheme. However, this solution only considers multimedia source rates and adopts a heuristic rate-compensation algorithm that cannot optimally determine the required congestion window size.

In summary, a media-aware congestion control mechanism that optimally determines the required congestion window size to maximize the long-term multimedia quality in a look-ahead (foresighted) rather than myopic manner is still missing.

*B. Contribution of our solution and paper organization*

In this paper, to overcome the abovementioned limitations, the proposed media-TCP aims to make the following contributions:

**1) Quality-centric packet-based congestion control**. The proposed media-TCP congestion control is quality-centric, meaning that it aims specifically at maximizing the received multimedia quality. Instead of applying a flow-based multimedia model, our solution takes into account the distortion impact and delay deadline of each packet, as well as the packets' interdependencies using a directed acyclic graph (DAG) [17]. Importantly, instead of reactively adapting the throughput, the proposed media-TCP actively and jointly optimizes the congestion window size as well as the transmission scheduling to provide differential services for different packet classes. Performing this joint optimization is very important in order to maximize the multimedia quality, because the optimal congestion window size depends on the transmission order of the multimedia packets in the transmission buffer.

**2) Foresighted adaptation using a Markov decision process framework.** We formulate the congestion control problem using a Finite-Horizon Markov decision process (FHMDP) framework in order to maximize the expected long-term multimedia quality, under a long-term TCP-friendliness constraint over the subsequent $K$ time slots (i.e. horizon-$K$ TCP-friendliness). Such foresighted planning is essential for multimedia streaming since it can consider, predict, and exploit the dynamic characteristics of the multimedia traffic in order to optimize the application performance over dynamic IP networks. We show that the complex FHMDP formulation can be decomposed into multiple optimal stopping problems [23]. Based on the structural results obtained from the decomposition, we present low-complexity threshold-based algorithms when the multimedia packets are coded either independently or interdependently.

**3) Quality-based fairness among coexisting streams.** Preserving the fairness among the coexisting streams represents an important issue [1][6]. However, even though the throughput/bandwidth is equally shared by the users, multimedia users can still experience very different qualities since various applications and source data may result in different traffic characteristics. Hence, instead of the throughput-based fairness proposed by most existing congestion control solutions, we focus in this paper on quality-based fairness. We adopt Jain's fairness index [26] on the multimedia qualities and show that the proposed media-TCP is able to achieve quality fairness among multimedia users. In [25], the authors also proposed a frame dropping scheme for min-max distortion fairness. However, the frame dropping approach is determined myopically, without considering the resulting TCP-friendliness to other flows.

In Table I, we compare the features of the proposed media-TCP with the existing TCP-friendly congestion control solutions for multimedia streaming.

TABLE I. COMPARISONS OF CURRENT CONGESTION CONTROL SOLUTIONS FOR MULTIMEDIA STREAMING.

| | Name of the adopted congestion control | Type of TCP-Friendliness | Multimedia support | Distortion impact consideration | Delay deadline/ Content dependency | Decision type |
|---|---|---|---|---|---|---|
| Towsley 2008 [5] | TCP-streaming | TCP | Playback buffering | No | No | Myopic |
| Bohacek 2003 [6] | TCP | TCP | Playback buffering | No | No | Myopic |
| Rejaie 1999 [13] | RAP | AIMD-based | Source rate adaptation – layered encoding | No | No | Myopic |
| Mark 2005 [14] | DTAIMD | AIMD-based | Optimal source rate is bounded due to buffer underflow at the receiver | No | No | Myopic |
| Seferoglu 2009 [10] | TFRC/FEC | Model-based | Application layer FEC | Yes | No | Myopic |
| Zakhor 1999 [8] | TFRC | Model-based | Source rate adaptation – packet size adaptation | Yes | No | Myopic |
| Zhang 2001 [11] | MSTFP | Model-based | Source rate adaptation – distortion minimization s.t. rate budget | Yes | No | Myopic |
| Our approach | Media-TCP | Model-based | Quality-centric congestion control | Yes | Yes | Foresighted |

The paper is organized as follows. In Section II, we first formulate the packet-based media-TCP congestion control problem for one media-TCP user. In Section III, we present the FHMDP framework used by the media-TCP user to determine the optimal transmission scheduling and congestion window size. In Section IV, we investigate how to decompose the FHMDP problem and provide structural results for solving this problem in different transmission scenarios. In Section V, we investigate multiple media-TCP users interacting in the same network with regular TCP users and discuss the quality-based fairness among the multimedia users. Simulation results are shown and discussed in Section VI, and Section VII concludes the paper.

## II. MEDIA-TCP CONGESTION CONTROL PROBLEM FOR ONE MEDIA-TCP USER

### A. Transport layer model

As in TCP, each packet transmission is acknowledged after a round-trip time (RTT) $Rtt$. Packet loss rate $p$ can be measured based on the packets' acknowledgements. We assume a model-based congestion control as in [4][8][11] that adapts the congestion window size to a long-term available TCP throughput calculated from the packet loss rate $p$ and the round-trip time $Rtt$ (the adaptation will be discussed in Section II.D). Let $l$ represesent the packet size. The long-term available TCP throughput can be approximated by $R_{AIMD}(a,b,Rtt,p) = \frac{l\sqrt{(2-b)a}}{Rtt\sqrt{2bp}}$ (bits/sec) [14], where the TCP congestion control is modeled as a special case of generic Additive Increase Multiplicative Decrease (AIMD) based congestion control with parameters $(a,b)$ [1] [14]. By substituting $(a,b)$ to $(1,0.5)$, we have the well-known TCP response function[2] $R_{TCP}(Rtt,p) = \frac{l}{Rtt}\sqrt{\frac{3}{2p}}$ (bits/sec). We assume a time-slotted system and set the time slot duration $T$ as $Rtt$ [3]. Denote the measured packet loss rate in time slot $k$ as $p^k$. We define the *expected TCP window size* in time slot $k$ as $W_{TCP}^k(p^k) = R_{TCP}(Rtt,p^k)\frac{Rtt}{l} = \sqrt{\frac{3}{2p^k}}$ (pkts/time slot), which can be viewed as a metric describing the network congestion in time slot $k$.

### B. Application layer multimedia model

Multimedia data is encoded and packetized into multiple data units at the application layer. A data unit usually encapsulates a video slice, which contains a set of macroblocks or an entire video frame (see e.g. the H.264 standard [21] for example). We assume that these data units will be packetized into RTP packets with the same size $l$ for transmission at the transport layer [4][14]. We also assume that these packets are classified into $M$ multimedia classes $\{CL_1,...,CL_M\}$. These packets are queued in different transmission (post-encoding) buffers for transmission. A class $CL_m$ in time slot $k$ is characterized by the set of parameters $\psi_m^k = \{N_m^k, N_m^{A,k}, N_m^{D,k}, Q_m, depth_m^k\}$. These parameters are discussed next.

(a) *Packet number:* Let $N_m^k$ represent the number of packets in the transmission buffer of class $CL_m$ in time slot $k$.

(b) *Arrival rate and discard rate:* Let $N_m^{A,k}$ denote the arrival rate, which represents the number of packets in class $CL_m$ that arrive in time slot $k$. Let $N_m^{D,k}$ denote the discard rate, which represents the number of packets in class $CL_m$ whose delay deadline expires in time slot $k$. A packet is purged from the buffer if 1) it is successfully transmitted or 2) its delay deadline is expired. Based on the packet arrivals and departures, the number of packets $N_m^k$ varies over time. In practice, if the multimedia data is pre-encoded, the arrival rate and discard rate in each time slot can be computed a priori[4]. In the case of real-time multimedia transmission, e.g. video conferencing, the arrival time and the delay deadline can be stochastically modeled [18].

---

[1] Given the current congestion window size $W$, the user increases its congestion window size as $W + a$ per RTT when there is no packet loss. When a packet loss event occurs, the TCP congestion control decreases the congestion widow size as $(1-b)W$ and retransmits the packets. A packet loss event can be either a timeout or receiving three duplicated acknowledgements in a row [14].

[2] A more sophisticated TCP response function can be obtained by Markov chain modeling [2], which considers the timeout duration.

[3] For simplicity, we assume a fixed RTT (time slot) in this paper. However, the proposed approach can be easily extended in a time-varying RTT environment.

(c) *Distortion impact:* We assume an additive distortion reduction for the packets similar to the one employed in [17][18]. Let the distortion reduction when the packets in class $CL_m$ are received and decoded at the receiver be $N_m^k Q_m$, where $Q_m$ represents the distortion impact of the class $CL_m$.

(d) *Depth*: Some classes of packets need to be received before others. Such interdependencies among the multimedia classes can be represented using a DAG [17]. Figure 1 gives an example of a DAG, in which MPEG video frames are classified into classes. More examples can be found in [17][18], and our solution is not restricted to any packet classification methods. Based on the DAG, if there is a path from class $CL_m$ to $CL_n$, we say the class $CL_m$ is an ancestor of $CL_n$, and $CL_n$ is a descendent of $CL_m$. Denote $\mathbf{Anc}_m$ and $\mathbf{Des}_m$ as the ancestor set and descendent set of the class $CL_m$. Let $depth_m^k$ represent the depth (the maximum distance) from class $CL_m$ to the root in the DAG in time slot $k$. For classes at the root of the DAG, we define its depth to be 0. Depth captures the importance of a class in terms of interdependency, which depends on the depths of the ancestor classes, i.e. $depth_m = \max_{CL_n \in \mathbf{Anc}_m} depth_n + 1$.

The DAG structure varies over time as a traveling tree in [18]. Figure 1 also shows the variation of the DAG when the packets in class $CL_1$ and $CL_3$ are transmitted in time slot $k$.

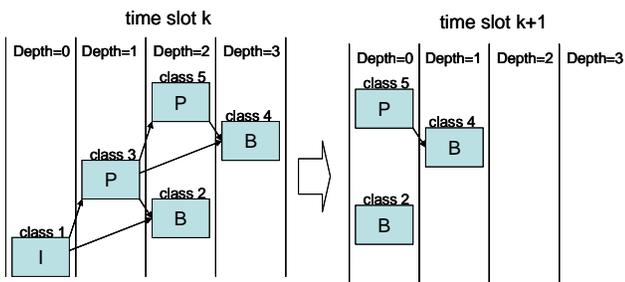

Fig. 1 Travelling tree example with MPEG IBPBP video frames.

Let $\pi_m^k \in \{1,0\}$ represent the transmission permission for transmitting the packets in class $CL_m$ in time slot $k$. We assume that if $\pi_m^k = 1$, then all of the $N_m^k$ packets of class $CL_m$ are transmitted in time slot $k$; if $\pi_m^k = 0$, then no packets in $CL_m$ are transmitted[5]. Denote $r_m^k(\pi_m^k) = N_m^k \pi_m^k l$ as the source rate of class $CL_m$ in time slot $k$, and denote $\mathbf{r}^k = [r_m^k, m=1,...,M]$ as the vector of rates for all the classes in time slot $k$. In addition, let $\boldsymbol{\pi}_{APP}^k = [\pi_m^i, i=1,...,k, m=1,...,M]$ represent the transmission permissions of all the classes from time slot 1 to time slot $k$. Hence, the availability $\rho_m^k$ of the class $CL_m$ at the receiver in time slot $k$ can be computed by $\rho_m^k(\boldsymbol{\pi}_{APP}^k) = I(\sum_{i=1}^k \pi_m^i \geq 1)$, where $I(\cdot)$ represents an indicator function. Based on the DAG, the actual distortion reduction for a class depends on whether or not its ancestors are available at the receiver. Hence, the actual distortion reduction of class $CL_m$ can be written as $Q_m^{act}(\boldsymbol{\pi}_{APP}^k) = Q_m \prod_{CL_n \in \mathbf{Anc}_m} \rho_n^k(\boldsymbol{\pi}_{APP}^k)$ and the resulting multimedia distortion reduction in time slot $k$ can be represented by:

$$\bar{Q}^k(\mathbf{r}^k(\boldsymbol{\pi}_{APP}^k)) = \sum_{m=1}^M Q_m^{act}(\boldsymbol{\pi}_{APP}^k) r_m^k(\pi_m^k). \quad (1)$$

### C. Conventional flow-based solutions

Most existing TCP-friendly congestion control solutions for multimedia streaming reactively adopt the available TCP throughput as a rate-budget constraint and maximize the immediate multimedia quality at the application layer (e.g. the rate-distortion optimization in [11] and the packet size adaptation in [7]). These solutions can be formulated using the following flow-based optimization.

**Myopic-Flow-Based Optimization[6]:**

$$\begin{aligned} &\underset{[\pi_m^k, m=1,...,M]}{\text{maximize}} \quad \bar{Q}^k([\pi_m^k, m=1,...,M]) \\ &\text{s. t.} \sum_{m=1}^M r_m^k \leq R_{TCP}(Rtt, p^k) \end{aligned} \quad (2)$$

Note that these solutions passively adapt the available TCP throughput $R_{TCP}(Rtt, p^k)$ to the network condition $p^k$ in each time slot. In contrast, we aim to propose a congestion control mechanism that adapts the congestion window size to both the network congestion and the specific characteristics and requirements of multimedia applications. Instead of shaping the traffic at the application layer to match the available throughput $R_{TCP}(Rtt, p^k)$, the proposed media-TCP jointly optimizes the congestion window size $W^k$ and the transmission permissions $\pi_m^k$ of classes at the transport layer to maximize the expected multimedia quality. In the next subsection, we discuss the media-TCP congestion control problem in more details.

---

[4] Assuming that the packets in class $CL_m$ has an arrival time $t_m$ (the time when the packets are ready for transmission in the buffer) and a delay deadline $d_m$, the arrival rate can then be calculated by $N_m^{A,k} = N_m^0 I(kT \leq t_m \leq (k+1)T)$, where $N_m^0$ represents the initial size of the class. The discard rate can be written as $N_m^{D,k} = N_m^k I(kT \leq d_m \leq (k+1)T)$.

[5] For illustration simplicity, we consider only binary transmission permission $\pi_m^k$ in this paper to transmit the entire class or not. However, similar approach can be applied to transmit partial data of a class by considering $\pi_m^k \in [0,1]$.

[6] We assume that the information of actual distortion reduction $Q_m^{act}(\boldsymbol{\pi}_{APP}^k)$ is available in both equation (2) and (4). Moreover, because of the retransmission error control in the transport layer, we ignore the impact of packet loss in the distortion reduction model.

### D. Proposed packet-based media-TCP solution

The proposed media-TCP congestion control is illustrated in Figure 2.

**At the application layer:** Multimedia RTP packets are classified into $M$ classes based on their interdependencies. Sophisticated packet classification can be performed in the application layer based on the different video coding structures. Based on the packet classification, the attributes $[\psi_m^k, m = 1,...,M]$ for each class can be determined as in [18][19][20].

**At the transport layer:** Media-TCP adopts the same error control as TCP, which retransmits the lost packets based on negative acknowledgements. However, unlike TCP which keeps retransmitting the lost packets until success, media-TCP will drop all the expired packets in the transmission buffer. Moreover, unlike TCP that adopts an AIMD-based congestion control, media-TCP adjusts the congestion window size relying on the following two components:

(a) *Transmission scheduler:* The transmission scheduler selects the classes of packets to transmit and also determines the number of packets to be sent in time slot $k$. Specifically, the packet scheduler computes the priority metrics $\mathbf{x}^k = [PM_1^k,...,PM_M^k] \in \mathbb{R}^M$ for all the classes to capture the marginal benefit (in terms of decreasing the expected distortion) when the packets in class $CL_m$ are transmitted in time slot $k$. In Section IV, we will discuss how to optimally determine these priority metrics based on the application attributes $[\psi_m^k, m = 1,...,M]$ and the network conditions. Then, the transmission scheduler sends the packets in the classes with positive priority metrics [7]. In other words, the transmission permission $\pi_m^k$ for each class and the resulting congestion window size $W^k$ can be determined as:

$$\pi_m^k(\mathbf{x}^k) = I(PM_m^k > 0), W^k(\mathbf{x}^k) = \sum_{m=1}^M N_m^k \pi_m^k(\mathbf{x}^k). \quad (3)$$

(b) *Network estimator:* The network estimator updates the packet loss rate[8] $p^k$ and evaluates the network congestion metrics in each time slot as in TCP. For example, a simple updating rule of the packet loss rate can be written as [11]: $p^{k+1}(p^k, W^k) = \alpha\, p^k + (1-\alpha)\hat{p}^k(W^k)$, where $\hat{p}^k(W^k)$ represents the realization of the packet loss rate in time slot $k$, and $\alpha$ represents the updating rates. In this paper, we focus on the joint optimization of the congestion window size and transmission scheduling by the transmission scheduler. For exposition simplicity, we assume that the only network congestion metric is the expected TCP window size $W_{TCP}^k(p^k)$. Note that other congestion metrics and more sophisticated updating rules (as in [4][11]) can be easily integrated into the proposed media-TCP.

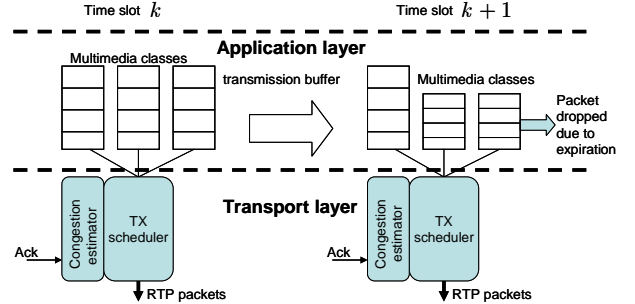

Fig. 2 System diagram of media-TCP in time slot $k$ and $k+1$.

We assume that the proposed media-TCP congestion control adheres to the following TCP-friendliness constraint.

**Definition 1 Horizon-$K$ TCP-friendliness:** A congestion control scheme is *horizon-$K$ TCP-friendly*, if and only if the congestion control window sizes from time slot $i$ to time slot $i + K - 1$ satisfy the condition $\sum_{k=i}^{i+K-1} \dfrac{W^k}{W_{TCP}^k(p^k)} \leq K$.

The horizon-$K$ TCP-friendliness constraint keeps the average TCP-friendliness ratio $W^k / W_{TCP}^k$ close to 1 over the horizon $K$. Based on this definition, in time slot $i$, the proposed media-TCP congestion control solves the following packet-based optimization.

**Foresighted-Packet-Based Optimization:**

$$\begin{aligned}
&\underset{\mathbf{x}^i \in \mathbb{R}^M}{\text{maximize}} \sum_{k=i}^{i+K-1} E[\overline{Q}^k([\pi_m^k(\mathbf{x}^k), m=1,...,M])] \\
&\text{s.t. } \frac{1}{K}\sum_{k=i}^{i+K-1} E\left[\frac{W^k(\mathbf{x}^k)}{W_{TCP}^k(p^k)}\right] \leq 1,\ 0 \leq W^k(\mathbf{x}^k) \leq W^{\max}
\end{aligned} \quad (4)$$

where $E[\cdot]$ represents the expected value and $W^{\max}$ represents the maximum congestion window size.

Comparing our media-TCP using the foresighted-packet-based optimization in equation (4) with the conventional solutions using the myopic-flow-based optimization in equation (2), the differences are:

1) Conventional solutions passively adapt the congestion

---

[7] A positive priority metric for a class indicates that the benefit (i.e. the resulting expected distortion reduction) of transmitting the packets in that class is greater than the cost of transmitting the packets. Hence, transmitting the packets of that class increases the total utility.

[8] Here, we assume that $Rtt$ remains constant. In practice, the round-trip time can also be updated using a similar updating rule. Our Theorems and Lemmas still hold in such cases.

window size to the network condition (e.g. the expected packet loss rate $p^k$). Our proposed media-TCP solution takes one step further by adapting the congestion window size to the network conditions as well as to the application characteristics by optimizing the transmission scheduling. This allows the user to adapt the congestion window size to the characteristics of the multimedia packets in the buffer to provide differential services for various packet classes. Hence, a packet class with a higher distortion impact or more stringent delay deadline has a higher chance to be transmitted, which is desirable for maximizing the received multimedia quality.

2) Media-TCP maximizes the expected long-term distortion reduction over a horizon-$K$ instead of solely maximizing the immediate distortion reduction as in the conventional solutions. This is especially important for multimedia applications with content dependencies. For example, in the IBPBP frame structure in Figure 1, the media-TCP user may want to plan the congestion window sizes for transmitting I and P frames, instead of myopically determining window sizes for the B frames.

3) Instead of performing a constrained rate optimization myopically at every time slot, media-TCP adopts the horizon-$K$ TCP-friendliness constraint. Note that the horizon-$K$ TCP-friendliness becomes the traditional rate budget constraint in equation (2) when $K = 1$. The larger horizons provide long-term TCP-friendliness, which leads to more flexible window sizes and a better expected long-term quality. However, the short-term TCP-unfriendliness can be high (especially when network condition is good, i.e. $W_{TCP}^k$ is large) and needs to be compensated in the subsequent time slots [12].

In the next section, we discuss how the foresighted-packet-based optimization in equation (4) can be solved by applying an FHMDP framework.

### III. FINITE-HORIZON MARKOV DECISION PROCESS

In this section, we first formulate the media-TCP congestion control problem in equation (4) using an FHMDP with Markovian state transition. The complexity of the FHMDP can be high due to the large state space. Hence, in the next section, we will decompose the problem into simpler sub-problems having smaller state spaces. We model this problem as an FHMDP due to the following reasons:

1) In numerous multimedia applications, multimedia traffic can be described by Markov models (as in [24]).

2) TCP operations are commonly modeled by discrete-time finite-state Markov chains (see e.g. [2][5][15]). Hence, the average TCP window size can be described by Markovian models based on the states of all the users (or the aggregate states as in [9]) in the network.

The FHMDP framework can be defined by the tuple $\{\mathcal{A}, \mathcal{S}, P, u, \gamma, \lambda, K\}$. The various components of the framework are described next:

(a) *Action:* We denote the action of the FHMDP in time slot $k$ as $a^k = [\pi_m^k, m = 1,...,M] \in \mathcal{A} = \{0,1\}^M$.

(b) *State and state transition:* We denote the state of the FHMDP in time slot $k$ as $s^k = \{W_{TCP}^k, \mathbf{N}^k\} \in \mathcal{S}^{Net} \times \mathcal{S}^{App} = \mathcal{S}$, where the expected TCP window size $W_{TCP}^k$ represents the network state and the number of packets $\mathbf{N}^k = [N_m^k, m = 1,...,M]$ in all the packet classes represents the application state. Let $\mathcal{S}^{Net} = \{0,...,W^{\max}\}$ represent the state space of the network state $W_{TCP}^k$, and let $\mathcal{S}^{App} = \{0,...,N^{\max}\}^M$ represent the state space of the application state, where $W^{\max}$ represents the maximum number of the window size and $N^{\max}$ represents the maximum number of packets in a class. Let $P: \mathcal{S} \times \mathcal{A} \times \mathcal{S} \to [0,1]$ denote the state transition function, which can be described for the network state and application state as follows:

1) The network state transition is described by the state transition probabilities $P(W_{TCP}^{k+1} \mid W_{TCP}^k)$, which can be evaluated by estimating the next possible packet loss rate $p^{k+1}$ given the current feedback $p^k$ as in [9][15]. In general, the number of TCP users in the network is large enough such that the network state transition is not impacted by a single user's action.

2) The application state transition is described by the state transition probabilities $P(\mathbf{N}^{k+1} \mid \mathbf{N}^k, a^k)$. The number of packets in each class varies over time depending on the action $a^k$. Note that each class can have its own arrival rate per time slot $N_m^{A,k} \geq 0$ and its own discard rate per time slot $N_m^{D,k} \geq 0$. Therefore, the application state transition can be computed as:

$$N_m^{k+1} = N_m^k(1 - \pi_m^k) + N_m^{A,k} - N_m^{D,k}(1 - \pi_m^k)$$
$$= \underbrace{(N_m^k - N_m^{D,k})(1 - \pi_m^k)}_{\text{remaining number of packets after packet departure}} + \underbrace{N_m^{A,k}}_{\text{packet arrivals}}. \quad (5)$$

This framework can be applied to both pre-encoded and real-time multimedia applications. For pre-encoded multimedia applications, the media-TCP user knows the state transitions for the entire multimedia session and solves the finite-horizon dynamic programming problem. For real-time multimedia applications, media-TCP can apply stochastic models to capture the state transitions of the applications [18].

Since the network state transition and the application state transition are independent, we define the overall state transition probabilities[9] as: $P(s^{k+1} \mid s^k, a^k) = P(W_{TCP}^{k+1} \mid W_{TCP}^k) P(\mathbf{N}^{k+1} \mid \mathbf{N}^k, a^k)$.

(c) *Utility, discount factor, and horizon definitions:*

First, we apply a positive Lagrangian multiplier $\lambda$ and modify equation (4) into an unconstrained optimization[10]:

$$\underset{a^i \in \mathcal{A}}{\text{maximize}} \sum_{k=i}^{i+K-1} \overline{Q}^k(a^k, s^k) - \lambda \left( \sum_{k=i}^{i+K-1} \frac{W^k}{W_{TCP}^k} - \sum_{k=i}^{i+K-1} 1 \right)$$

$$= \underset{a^i \in \mathcal{A}}{\text{maximize}} \sum_{k=i}^{i+K-1} u^k(a^k, s^k) \quad (6)$$

where $u^k(a^k, s^k) = \overline{Q}^k(a^k, s^k) - \lambda\left(W^k / W_{TCP}^k - 1\right)$ is referred to as the *instantaneous utility* in time slot $k$. The second term of the instantaneous utility can be interpreted as the TCP window size deviation cost. The Lagrangian multiplier $\lambda$ [11] determines how the media-TCP user favors the TCP-friendliness over the multimedia quality. Based on the unconstrained optimization, the objective of the FHMDP in time slot $i$ is defined as:

$$\chi^i(s^i) = \arg\max_{a \in \mathcal{A}} \sum_{k=i}^{i+K-1} \gamma^{k-i} u^k(s^k, a^k), \quad (7)$$

where $\chi^i(s^i)$ represents the optimal congestion control policy given the state $s^i$ at the time slot $i$, and $\gamma$ represents the discount factor ($0 \le \gamma \le 1$). Note that equation (7) is equivalent to the unconstrained optimization in equation (6) when $\gamma = 1$. Since the Markovian models of the network state and application state transition may not be accurate, the discount factor $\gamma$ is set smaller than 1 to alleviate the impact of the inaccurate future utilities. The tradeoff between the TCP-friendliness and multimedia quality with different $\lambda$ and $\gamma$ will be discussed in Section VI.

Note that the media-TCP not only maximizes the instantaneous utility, but also the expected future utilities, which are expressed using the expected utility-to-go, which is defined next.

**Definition 2. Expected utility-to-go:** Define the expected utility-to-go at the last time slot of the horizon as $J_\mu^{i+K-1}(s) = u^{i+K-1}(s, \mu(s)), \forall s \in \mathcal{S}$, where $\mu : \mathcal{S} \to \mathcal{A}$ represents a stationary mapping from the given state to an action. We define the expected utility-to-go in time slot $k = i, \ldots, i+K-2$ as:

$$J_\mu^k(s^k) = u^k(s^k, \mu(s^k)) + \gamma \sum_{s^{k+1} \in \mathcal{S}} P(s^{k+1} \mid s^k, \mu(s^k)) J_\mu^{k+1}(s^{k+1}).$$

The optimal congestion control policy in time slot $i$ can then be rewritten as:

$$\chi^i(s^i) = \arg\max_\mu J_\mu^i(s^i). \quad (8)$$

The system diagram of the proposed media-TCP congestion control using FHMDP framework is shown in Figure 3. The media-TCP user repeats the following steps at each time slot $i$:

1) Calculate the application state and network state transition probabilities.
2) Evaluate the expected utility-to-go at the various time slots of the entire horizon, i.e. from $J_\mu^{i+K-1}(s^{i+K-1}), \forall s^{i+K-1} \in \mathcal{S}$ to $J_\mu^{i+1}(s^{i+1}), \forall s^{i+1} \in \mathcal{S}$.
3) Based on the expected utility-to-go, the user updates the policy using

$$\mu(s) \leftarrow \arg\max_{a \in \mathcal{A}} \left\{ u(s^i, a) + \gamma \sum_{s^{k+1} \in \mathcal{S}} P(s^{k+1} \mid s^k, a) J_\mu^{k+1}(s^{k+1}) \right\}, (9)$$

and obtains the optimal action given the current state $s^i$.

The approach is similar to the "receding horizon control" in the optimal control literature [15]. Note that the complexity of solving the FHMDP directly is extremely high: it is proportional to the square of the number of states, and the number of states is exponential in the number of classes, i.e. $(N^{\max})^M$. Hence, it is important to decompose the FHMDP problem into sub-problems with smaller state space to reduce the complexity of solving the problem.

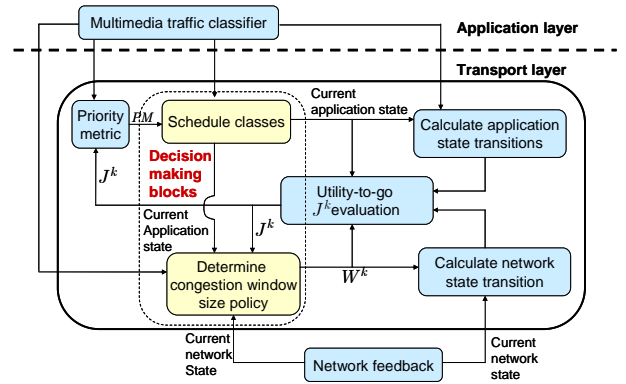

Fig. 3 System diagram of the proposed media-TCP congestion control.

## IV. STRUCTURAL SOLUTIONS OF MEDIA-TCP CONGESTION CONTROL

In this section, we decompose the FHMDP problem in Section III. We derive structural results that provide the optimal values of the priority metrics $\mathbf{x}^{i^*}$ for the media-TCP congestion control problem described in Section II.D to facilitate a low-complexity threshold-based congestion control as shown in equation (3). In Section IV.A, we discuss the decomposition of the FHMDP when the applications are coded independently. Then, in Section IV.B, we take the interdependencies among packets into account and discuss how to derive the structural results in this case.

---

[9] In this work, we assume that the state transition probabilities can be calculated a priori. In fact, the state transition probabilities can be learned on the fly as long as the transitions of the states are semi-stationary.

[10] To simplify notation, we ignore the expectation $E[\bullet]$ in our FHMDP formulation.

[11] $\lambda$ can be chosen based on the dual problem of equation (4) with the constraint $\sum_{m=1}^M \left( Q_m - \lambda / W_{TCP}^k \right) N_m^k = 0$, which suggests $\lambda = \tilde{W}_{TCP} \tilde{Q}_m$, where $\tilde{Q}_m$ represents the average distortion reduction per packets in class $CL_m$, and $\tilde{W}_{TCP}$ is the time average $W_{TCP}^k$ over the horizon.

*A. Decomposition with independently coded packets*

We first discuss the case when all the packets are independently coded (for example, video streams are coded using motion-JPEG), i.e. all the classes have the same depth, $depth_m = 0, m = 1,...,M$. We first examine the structure of $J_\mu^k(s^k)$ and define the following property of the expected utility-to-go.

**Definition 3. Separable expected utility-to-go:** The expected utility-to-go $J_\mu^k(s^k)$ in time slot $k$ is separable if and only if it can be written in the form of $J_\mu^k(s^k) = \sum_{m=1}^M J_{\mu,m}^k(W_{TCP}^k, N_m^k) + C(W_{TCP}^k)$, where $J_{\mu,m}^k(W_{TCP}^k, N_m^k)$ represents the utility-to-go component of a specific class $CL_m$, and $C(W_{TCP}^k)$ represents a term that only depends on the network state $W_{TCP}^k$.

Next, we show that the expected utility-to-go is separable for multimedia applications with independently-coded packets.

**Theorem 1:** The expected utility-to-go $J_\mu^k(s^k), \forall s^k \in S, k = i,...,i+K-1$ over the horizon are separable, and the utility-to-go component $J_{\mu,m}^k(W_{TCP}^k, N_m^k)$ of class $CL_m$ is a nondecreasing function of $N_m^k$.

*Proof:* See Appendix B.

Based on Theorem 1, we have the following remarks:

*Remark 1:* The separation of the expected utility-to-go $J_\mu^k(s^k)$ in Theorem 1 suggests that the utility-to-go evaluation can be decomposed into $M$ independent FHMDP sub-problems. Each sub-problem computes a utility-to-go component $J_{\mu,m}^k(W_{TCP}^k, N_m^k)$ for a class (see equation (17) for the optimization sub-problems). The decomposition significantly reduces the overall complexity originally proportional to $(N^{\max})^M$ to a complexity proportional to $MN^{\max}$.

*Remark 2:* The nondecreasing property of $J_{\mu,m}^k(W_{TCP}^k, N_m^k)$ in $N_m^k$ in Theorem 1 shows that the more packets in the transmission buffer of class $CL_m$, the media-TCP user has higher expected utility for the specific class when the application is independently coded. Next, we investigate how the user determines the optimal congestion control policy when it knows the number of packets in the transmission buffer of each class. The following theorem presents the structural results of solving the FHMDP problem.

**Theorem 2. Structural results of the proposed media-TCP with independent packets:** Given the state $s^i$ in time slot $i$, the optimal policy of the transmission permission for a class $CL_m$ is $\pi_m^{i*}(s^i) = I(PM_m^{i*}(s^i) > 0)$, where the optimal priority metric $PM_m^{i*}(s^i)$ is computed by:

$$PM_m^{i*}(s^i) = \left(Q_m - \frac{\lambda}{W_{TCP}^i}\right)N_m^i + \gamma \sum_{s^{i+1} \in \mathcal{S}} P(W_{TCP}^{i+1} \mid W_{TCP}^i) \begin{pmatrix} J_{\mu,m}^{i+1}(W_{TCP}^{i+1}, N_m^{A,i}) \\ -J_{\mu,m}^{i+1}(W_{TCP}^{i+1}, N_m^i - (N_m^{D,i} - N_m^{A,i})) \end{pmatrix}$$

(10)

*Proof:* See Appendix C.

Theorem 2 indicates that when the packets are coded independently, the media-TCP congestion control problem becomes an optimal stopping problem [23], where the media-TCP user transmits the packets of a certain class if and only if the priority metric of the class is positive. The priority metrics $\mathbf{x}^{i*}(s^i) = [PM_m^{i*}(s^i), m = 1,...,M]$ quantify the benefits of transmitting packets from various classes as opposed to not transmitting them in time slot $i$. Importantly, in addition to the distortion impact $Q_m$, the media-TCP user needs to consider the arrival rates and discard rates of the various classes. If a class $CL_m$ has numerous expiring packets (i.e. $N_m^{D,i} - N_m^{A,i}$ is large), it can be shown that the respective class has a larger $PM_m^{i*}(s^i)$ to be transmitted in time slot $i$, instead of waiting for a future time slot. Moreover, it can be shown that at a time slot with a better network state (larger $W_{TCP}^i$), we have larger priority metrics from equation (10) for all the classes and hence, more classes are able to obtain the transmission permissions. Based on Theorem 2, we have the following remarks:

*Remark 3:* In equation (10), as $\gamma$ approaches 0, the user prefers to prioritize the packet classes based on their distortion impact values $Q_m$. As $\gamma$ approaches 1, the user increasingly weights the impact from the arrival rate and discard rate on the future expected utility. Note that when $\gamma = 0$, the FHMDP problem in equation (7) becomes a myopic optimization that merely optimize the instantaneous utility, which is equivalent to solving an unconstrained optimization of the conventional solution in equation (2).

*Remark 4:* Note that the optimal congestion control policy $\chi^i(s^i)$ in equation (7) includes the optimal priority metrics $\mathbf{x}^{i*}(s^i)$ and the congestion window size $W^{i*}(s^i)$. Theorem 2 provides the optimal priority metrics $\mathbf{x}^{i*}(s^i) = [PM_m^{i*}(s^i), m = 1,...,M]$. Based on this, the optimal congestion window size $W^{i*}(s^i)$ of the media-TCP user can be written as $W^{i*}(s^i) = \sum_{m=1}^M N_m^i I(PM_m^{i*}(s^i) > 0)$ and the resulting expected multimedia quality can be computed by $\overline{Q}^i(s^i) = \sum_{m=1}^{M} Q_m N_m^i I(PM_m^{i*}(s^i) > 0)$. Note that the optimal policy varies with both the application state and the network state in time slot $i$ ($s^i = \{W_{TCP}^i, \mathbf{N}^i\}$).

In Appendix A, Algorithm 1 provides the specific procedures for computing the optimal congestion policy $\chi^i(s^i)$ when the packets are independent. The time complexity of the algorithm is $O(KM(W^{\max}N^{\max})^2)$.

*B. Decomposition with interdependently coded packets*

In this subsection, we investigate the decomposition of FHMDP problem when the packets have interdependencies, described by a DAG as introduced in Section II.A. The following theorem presents the structural results of solving the FHMDP problem.

**Theorem 3. Structural results of media-TCP with interdependent packets:** Given the DAG and the state $s^i$ in time slot $i$, the FHMDP problem can be solved by repeating the following two phases:

*Phase 1*. Select packet classes to transmit at the current time slot at the depth $depth_n^i = j - 1$:
$$\pi_m^{i*}(s^i) = I(PM_m^{i*}(s^i) > 0), \forall CL_m \in \{CL_n, depth_n^i = j - 1\},$$
where
$$PM_m^{i*}(s^i) = \left(Q_m^{act} - \frac{\lambda}{W_{TCP}^i}\right) N_m^i + \gamma \sum_{s^{i+1} \in \mathcal{S}} P(W_{TCP}^{i+1} \mid W_{TCP}^i) \begin{pmatrix} J_{\mu,m}^{i+1}(W_{TCP}^{i+1}, N_m^{A,i}) - \\ J_{\mu,m}^{i+1}(W_{TCP}^{i+1}, N_m^i - (N_m^{D,i} - N_m^{A,i})) \end{pmatrix}, (11)$$
and $j$ represents the number of iterations.

*Phase 2*. Update the actual distortion impact of each class: $Q_m^{act} = Q_m \prod_{\forall CL_n \in \mathbf{Anc}_m^i} \pi_n^{i*}$, $m = 1, ..., M$.

*Proof:* The proof can be easily provided by considering the DAG structure. We omit the proof due to space limitations.

In Phase 1, the media-TCP user selects packet classes for transmission by applying Theorem 3 starting from the classes at the root of the DAG. Since classes with the same depth are independent of each other, Theorem 2 can be applied to Phase 1 for classes with the same depth. Phase 2 indicates that if a class has no transmission permission, i.e. $\pi_m^{i*}(s^i) = 0$, the media-TCP user set all its descendents' distortion impact to 0, i.e. $Q_n^{act} = 0$ for $\forall CL_n \in \mathbf{Des}_m$ (see Section II.B). Based on the DAG, since the distortion impact of a class is only influenced by the ancestors, the greedy algorithm in Theorem 3 starting from the root provides the optimal congestion policy. The two phases are repeated until the maximum depth of the DAG is reached.

In Appendix A, Algorithm 2 provides the procedures for computing the optimal congestion control policy $\chi^i(s^i) = \{\mathbf{x}^{i*}(s^i), W^{i*}(s^i)\}$ when the packets are interdependently coded. Assuming that the DAG has the maximum depth $\overline{D}$ and the number of classes per depth is $\overline{M}$ on average, the complexity of the algorithm can be represented by $O(\overline{M}\overline{D}K(W^{\max}N^{\max})^2)$.

## V. FAIRNESS AMONG MULTIPLE MEDIA-TCP STREAMS

In the previous sections, we focus on only one media-TCP user interacting with multiple regular TCP users in the same network. In this section, we assume that there are $V$ media-TCP users interacting with other TCP users in a network and investigate the competition among the multimedia users. Denote $\mathbf{V} = \{V_n, n = 1, ..., V\}$ as the set of the media-TCP users. Denote user $V_n$'s expected multimedia distortion reduction in time slot $k$ as $\overline{Q}_n^k$. We assume a saturated condition (i.e. all the users continuously have their source traffic fed into their transmission buffers). We apply the well-known Jain's fairness index [26] to quantify the fairness among the $V$ media-TCP users:

$$\mathcal{F}^k = \frac{\left(\sum_{n=1}^{V} \overline{Q}_n^k\right)^2}{V \sum_{n=1}^{V} \left(\overline{Q}_n^k\right)^2}. \quad (12)$$

The fairness index $\mathcal{F}^k$ measures the quality deviation of the multimedia applications. It varies as the media-TCP users make their own decisions at each time slot. Note that the index is always bounded by 1. The quality-based fairness is reached, i.e. $\mathcal{F} = 1$, only when all the media-TCP users have the same multimedia quality.

Following the TCP response function introduced in Section II.B and the packet loss rate updating rules in Section II.D, the expected network state of a user $V_n$ in the next time slot can be expressed as $E[W_{TCP,n}^{k+1}(p_n^k, W_n^k)] = \sqrt{\dfrac{1.5}{p_n^{k+1}(p_n^k, W_n^k)}}$.

Similarly, we denote $E[PM_{mn}^{k+1}]$ as the expected priority metric of $CL_{mn}$ (the $m$-th class of user $V_n$) in time slot $k+1$, which can be shown as a function of $W_n^k$. Then, we can prove the following lemma.

**Lemma 1:** For user $V_n$, its priority metrics $\{E[PM_{mn}^{k+1}(W_n^k)], \forall CL_{mn} \in V_n\}$ are all nonincreasing functions of $W_n^k$, if
1) $\gamma = 0$, or
2) $0 < \gamma \leq 1$, $N_m^{D,k} = N_m^k$, $\forall CL_{mn} \in V_n, k = 1, ..., K$

*Proof:* For both conditions, the priority metric of class $CL_{mn}$ can be written by $E[PM_{mn}^{k+1}(W_n^k)] = Q_{mn} - \lambda / E[W_{TCP}^{k+1}(p_n^k, W_n^k)]$ based on equation (10). Since the estimated packet loss rate $p^{k+1}(p^k, W^k)$ is in general a monotonically nondecreasing function [12] of the congestion window size $W_n^k$ [11], it is straightforward that both the expected window size $E[W_{TCP}^{k+1}(p_n^k, W_n^k)]$ and the expected priority metrics $E[PM_{mn}^{k+1}(W_n^k)]$ in the next time slot are nonincreasing functions of $W_n^k$. ∎

---
[12] The packet loss rate can be modeled as an M/M/1/K queue at the bottleneck link that reacts to the summation of the window sizes of all the users.

Lemma 1 indicates that the priority metrics $E[PM_{mn}^{k+1}(W_n^k)], \forall CL_{mn} \in V_n$ are nonincreasing functions of $W_n^k$ when the users apply the myopic media-TCP or when the packets in the buffer expire in the next time slot. In these two cases, the priority metrics are dominated by the first term in equation (10). The second condition requires the delay deadlines of the classes to be stringent, which is more likely to be true in the case of real-time streaming, as opposed to the pre-encoded streaming applications. In these two cases, Lemma 1 indicates that the competition among users makes it impossible for a user to excessively increase its congestion window size in order to improve its own quality. The increase of the congestion window size $W_n^k$ may decrease the priority metrics and hence, reduce the resulting distortion reduction in the next time slot. Following Remark 4 in Section IV, the multimedia distortion reduction of user $V_n$ in the next time slot can be written as

$$\bar{Q}_n^{k+1}(W_n^k) = \sum_{\forall CL_{mn} \in V_n} Q_{mn} N_{mn} I(E[PM_{mn}^{k+1}(W_n^k)] > 0), \quad (13)$$

where $Q_{mn}$ represents the distortion impact of class $CL_{mn}$. Comparing $\bar{Q}_n^{k+1}(W_n^k)$ with the current multimedia distortion reduction $\bar{Q}_n^k$, the variation appears only for the classes whose priority metrics change sign. If we denote $\mathbf{M}_n^k$ as the set of classes whose priority metrics follow $PM_{mn}^k E[PM_{mn}^{k+1}] < 0$, we can rewrite equation (13) as

$$\bar{Q}_n^{k+1} = \begin{cases} \bar{Q}_n^k, & \text{if } \mathbf{M}_n^k = \varnothing \\ \bar{Q}_n^k + \Delta \bar{Q}_n^k, & \text{if } PM_{mn}^k < E[PM_{mn}^{k+1}], \forall CL_{mn} \in \mathbf{M}_n^k \\ \bar{Q}_n^k - \Delta \bar{Q}_n^k, & \text{if } PM_{mn}^k > E[PM_{mn}^{k+1}], \forall CL_{mn} \in \mathbf{M}_n^k \end{cases}, \quad (14)$$

where $\Delta \bar{Q}_n^k = |\bar{Q}_n^{k+1} - \bar{Q}_n^k| = \sum_{\forall CL_{mn} \in \mathbf{M}_n^k} Q_{mn} N_{mn} \geq 0$ represents the difference of the expected distortion reduction of user $V_n$ in time slot $k$. Based on equation (14), we next prove the sufficient condition for achieving the discussed quality-based fairness.

**Lemma 2:** The difference of the fairness index is nonnegative, i.e. $\Delta \mathcal{F}^k = \mathcal{F}^{k+1} - \mathcal{F}^k \geq 0$, if

$$\sum_{V_n \in \mathbf{V}} \left(\bar{Q}_n^k\right)^2 \sum_{V_n \in \mathbf{V}} \Delta \bar{Q}_n^k \geq \sum_{V_n \in \mathbf{V}} \bar{Q}_n^k \sum_{V_n \in \mathbf{V}} \bar{Q}_n^k \Delta \bar{Q}_n^k. \quad (15)$$

*Proof:* We omit the proof here due to space limitations. A similar proof can be found in [16].

Lemma 2 provides a sufficient condition that ensures a nondecreasing fairness index $\Delta \mathcal{F}^k$. Since the index is bounded by 1, the interaction among users asymptotically drives the fairness index to 1 [16]. Based on Lemma 2, the following Theorem provides the sufficient conditions for multiple myopic media-TCP users to reach the quality-based fairness.

**Theorem 4:** The fairness index of multiple media-TCP users $V_n \in \mathbf{V}$ converges to 1, i.e. $\mathcal{F}^\infty = 1$, if the following sufficient conditions are satisfied:
1) $Q_{mn} = Q_{mn'} = Q_m \; \forall V_n, V_{n'} \in \mathbf{V}$
2) $N_{mn} \geq N_{m'n'}$ for any $Q_m \geq Q_{m'}$, $\forall V_n, V_{n'} \in \mathbf{V}$.
3) $E[PM_{mn}^{k+1}(W_n^k)], \forall CL_{mn} \in V_n, \forall V_n \in \mathbf{V}$ are all nonincreasing functions of $W_n^k$ using the same $\lambda$.

*Proof:* See Appendix D.

The first condition in Theorem 4 indicates that the media-TCP users apply the same set of $[Q_m, m = 1,...,M]$ to classify their $M$ classes of the multimedia applications. The second condition indicates that for all the users in the network, users always have more packets in a class with higher distortion impact than a class with lower distortion impact. This condition is commonly seen in many video coding techniques. For example, in MPEG video frames, I-frames usually contain much more information bits than P-frames and B-frames. The third condition is discussed in Lemma 1. Based on equation (14), as long as the priority metrics $E[PM_{mn}^{k+1}(W_n^k)]$ are nonincreasing functions of $W_n^k$ for all the classes, we can show that a user with a larger multimedia quality $\bar{Q}_n$ will always have a smaller quality change $\Delta \bar{Q}_n$ when users applying media-TCP to change their congestion window size. In Appendix D, we prove that this allows the proposed media-TCP to satisfy the sufficient condition in Lemma 2 and hence the quality-based fairness index converges to 1. Finally, the three conditions in Theorem 4 lead to $\mathcal{F}^\infty = 1$ for media-TCP users $V_n \in \mathbf{V}$.

## VI. SIMULATION RESULTS

In this section, we simulate the proposed congestion control scheme using different video sequences: "Forman", "Mobile", and "Coastguard" (at a frame rate of 30 Hz, CIF format). The sequences are encoded using an embedded scalable video codec [20] at the bitrate of 1500Kbps. We assume that each Group of Picture (GOP) contains 16 frames and each of them can tolerate a playback delay of {133, 266, 400, 533} ms. We set the packet length up to 1000 bytes and the video packets are classified into sixteen classes based on their spatial and temporal interdependencies as in [19][20]. Table II provides a summary of the classifications of the sequences. We simulate the video transmission using MATLAB using the simulation settings in Figure 4. There are 20 regular TCP users and the resulting average RTT for the video packets is 133ms.

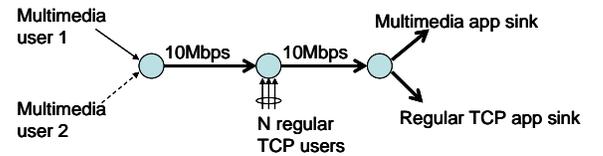

Fig. 4 The simulation settings.

*A. Tradeoff between multimedia performance and TCP friendliness*

First, we simulate the case without the media-TCP user 2. We focus on the media-TCP user 1 streaming the "Coastguard" sequence using the proposed media-TCP congestion control with different Lagrangian multipliers and discount factors. Based on the measured packet loss rate, the media-TCP user applies a Markov chain model (similar to the model applied in [9]) on the network states with an expected TCP window size $W_{TCP}^k = 16$ per RTT, and the horizon $K = 4$ RTT. Figure 5 shows the tradeoff between multimedia quality and the horizon-$K$ TCP-friendliness. Larger $\lambda$ provides better TCP-friendliness, but achieves lower multimedia quality, because the quality gain is weighed less than the cost within the instantaneous utility. The results also show that the foresighted approach with larger $\gamma$ significantly improves the multimedia quality while maintaining moderate TCP-friendliness. However, in this paper, we focus on deriving the optimal solution when the environment (i.e. the state transition probabilities) and the utility are perfectly known. If the transition probabilities are not perfect, a larger $\gamma$ can lead to a worse learning performance. The selection of $\lambda$ and $\gamma$ for media-TCP using online learning for the case when the environment is unknown represents an interesting future research direction.

TABLE II. CLASSIFICATION OF THE SEQUENCES.

| Class $CL_m$ | 1 | 2 | 3 | 4 | 5~8 | 9~16 |
|---|---|---|---|---|---|---|
| $Q_m$ (dB/pkt) range | 0.154 | 0.153 | 0.09 | 0.08 | ~0.072 | ~0.053 |
| $N_m^0$ of "Coastguard" per GOP | 17 | 17 | 12 | 12 | 5 | 4 |
| $N_m^0$ of "Forman" per GOP | 34 | 34 | 8 | 8 | 4 | 0 |
| $N_m^0$ of "Mobile" per GOP | 30 | 30 | 13 | 13 | 1 | 0 |

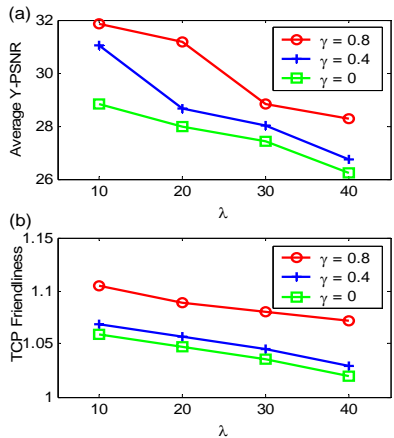

Fig. 5 (a) Average Y-PSNR of Coastguard sequence, (b) resulting TCP-friendliness versus different Lagrangian multipliers (playback delay: 266ms, $K = 4$ RTT).

### B. Comparisons against alternative congestion control solutions for multimedia applications

We simulate separately the streaming of "Coastguard" sequence as well as "Forman" sequence using three different approaches: 1) our proposed packet-based media-TCP congestion control (MT) that jointly optimizes the transmission scheduling and congestion window size; 2) a flow-based rate-distortion optimization approach (RD) [11] to optimize the transmission scheduling by adapting the sending rate to the available TCP throughput; 3) passive multimedia transmission directly over TCP connections (PA) as in [5].

Figure 6 shows the average video quality of various approaches using different playback delays and clearly demonstrates that the joint transmission scheduling and congestion control optimization is essential for real-time multimedia transmission. Our proposed approach significantly outperforms the others especially when the playback delay is smaller than 400ms (which is common in numerous real-time video streaming and videoconferencing applications), because it is able to jointly optimize the congestion window size as well as the transmission scheduling by considering the distortion impacts, delay deadlines, and interdependencies of the packets.

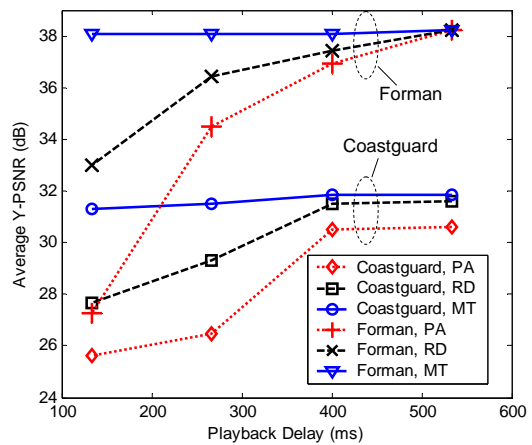

Fig. 6 Average received video quality using different TCP congestion control for multimedia transmission (For MT approach, $\lambda = 10, \gamma = 0.8$, $K = 4$ RTT).

### C. Fairness among multiple media-TCP streams

In this subsection, we validate the quality-based fairness among multiple users using media-TCP. We simulate the case when multimedia user 1 streams "Coastguard" sequence and multimedia user 2 streams "Mobile" sequence simultaneously using the same simulation settings in Figure 4 with 20 regular TCP users. The playback delay is set as 533ms. Figure 7 shows the congestion window size and the video quality over time when the multimedia users apply the passive multimedia transmission approach (PA) directly over TCP connections as in [5]. It is shown that although the congestion window sizes of the multimedia users follow the average TCP window size of the 20 regular TCP users, the video quality gap between the two sequences is always larger than 3 dB.

On the other hand, Figure 8 shows the congestion window size and the video quality over time when the multimedia users apply the proposed media-TCP congestion control (MT) algorithms. We classify the video packets of both sequences to satisfy the first two sufficient conditions in Theorem 4. We also set a small discount factor $\gamma = 0.1$ to ensure that the priority metrics $E[PM_{mn}^{k+1}(W_n^k)]$ are nonincreasing functions

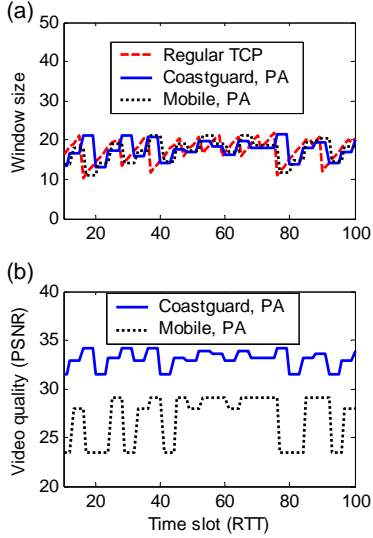

Fig. 7 (a) Congestion window sizes over time for the two multimedia users and the average congestion window size of the 20 regular TCP users. (b) Video quality over time for the two multimedia users using PA approach (playback delay: 533ms).

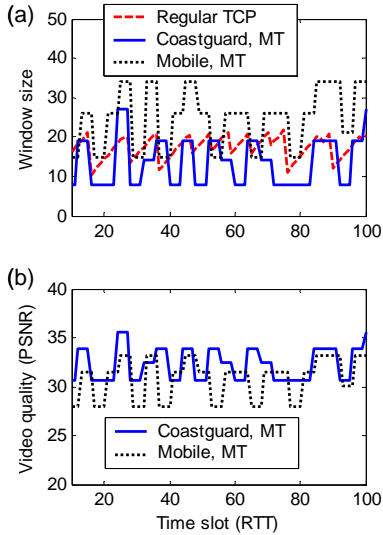

Fig. 8 (a) Congestion window sizes over time for the two multimedia users and the average congestion window size of the 20 regular TCP users. (b) Video quality over time for the two multimedia users using MT approach ($\lambda = 10$, $\gamma = 0.1$, $K = 4$ RTT, playback delay: 533ms).

of $W_n^k$ for all the classes. It is shown that media-TCP results in closer streaming qualities for the two multimedia users at the cost of moderate TCP unfriendliness. Hence, by applying the proposed media-TCP, the multimedia users fairly share the resources in terms of video quality.

We further increase the number of the regular TCP users using the same simulation settings. The resulting congestion window sizes and the video qualities are summarized in Table III. It is shown that the proposed media-TCP congestion control is able to maintain the streaming qualities for both sequences as the number of users in the network increases, while having limited impact on the other regular TCP users. This is because the media-TCP is able to better utilize the resource by prioritizing the video classes for transmission, in addition to merely adapting the congestion window size.

## VII. CONCLUSIONS

In this paper, we formulate a media-aware congestion control for multimedia transmission using FHMDP that explicitly considers the distortion impacts, delay deadlines and interdependencies of the various multimedia classes. The proposed approach not only adapts the congestion window size given the measured packet loss rate, but also optimally prioritizes the multimedia classes for transmission and hence, further improves the multimedia quality. We show that this complex FHMDP problem can be decomposed into simpler optimal stopping problems, thereby significantly reducing the complexity of solving the problem. The simulation results show that the proposed foresighted media-TCP significantly outperforms the conventional TCP-friendly congestion control schemes in terms of quality, especially for real-time streaming with a small playback delay. Moreover, unlike the conventional congestion control approaches focusing on the throughput-based fairness, our solution maintains the quality-based fairness among the multimedia users, which improves the overall streaming quality by utilizing the available bandwidth resources more efficiently.

TABLE III. THE COMPARISONS OF THE VIDEO QUALITIES USING THE PA APPROACH AND THE PROPOSED MT APPROACH WHEN THE NUMBER OF USERS IN THE NETWORK INCREASES.

| (playback delay: 533ms) | PA approach | | | | MT approach ($\lambda = 10$, $\gamma = 0.1$, $K = 4$ RTT) | | | |
|---|---|---|---|---|---|---|---|---|
| Number of TCP users | Avg. PSNR of user 1(dB) | Avg. PSNR of user 2 (dB) | Avg. PSNR of the users (dB) | Avg. window size of TCP users | Avg. PSNR of user 1(dB) | Avg. PSNR of user 2 (dB) | Avg. PSNR of the users (dB) | Avg. window size of TCP users |
| N = 20 | 32.07 | 27.46 | 29.76 | 17.00 | 31.58 | 30.43 | 31.00 | 16.95 |
| N = 25 | 31.14 | 25.61 | 28.37 | 14.14 | 31.43 | 30.40 | 30.91 | 13.80 |
| N = 30 | 29.88 | 23.38 | 26.63 | 11.70 | 30.29 | 30.22 | 30.25 | 11.51 |

## APPENDIX A

*Algorithm 1 Media-TCP congestion control with independent packets*

**For time slot** $k = i$, **given the current state** $s^i, \lambda, \gamma, K$
  **Set** $k = i + K - 1$;
  **While** $k \geq i$
    **For all classes** $CL_m$
      **Compute all** $J_m^k(W_{TCP}^k, N_m^k)$ **from equation** (17),
        **for** $N_m^k \in \{0,...,N_m^{\max}\}, W_{TCP}^k \in \{0,...,W_{TCP}^{\max}\}$;
      **Compute** $PM_m^{k*}(s^k)$ **and** $\pi_m^{k*}(s^k)$ **in equation** (10);
    **End for**
    **Set** $k = k - 1$;
  **End while**
  **Set** $PM_m^{i*} = PM_m^{k*}(s^i)$ **for** $m = 1,...,M$;
  **Set** $W^{i*} = \sum_{\forall m, PM_m^{i*} \geq 0} N_m^i$;

*Algorithm 2 Media-TCP congestion control with interdependent packets*

**For time slot** $k = i$, **given the current state** $s^i, \lambda, \gamma, K$
  Set $k = i + K - 1$;
  **While** $k \geq i$
    Set $W^k = 0$ and $depth = 0$;
    **For all classes** $CL_m$ **with** $depth_m^k = depth$
      **Compute all** $J_m^k(W_{TCP}^k, N_m^k)$ **from equation** (17),
        for $N_m^k \in \{0,...,N_m^{\max}\}, W_{TCP}^k \in \{0,...,W_{TCP}^{\max}\}$;
      **Compute** $PM_m^{k*}(s^k)$ **and** $\pi_m^{k*}(s^k)$ **in equation** (11);
      **If** $PM_m^{k*}(s^k) < 0$, **set** $Q_n = 0$ **for** $\forall CL_n \in \mathbf{Des}_m$;
      Set $depth = depth + 1$;
    **End for**
    Set $k = k - 1$;
  **End while**
  Set $PM_m^{i*} = PM_m^{k*}(s^i)$ for $m = 1,...,M$;
  Set $W^{i*} = \sum_{\forall m, PM_m^{i*} \geq 0} N_m^i$;

## APPENDIX B

*Proof of Theorem 1:* Without losing generality, we assume $i = 1$. Since packets are independent, the distortion reduction $\bar{Q}^k(a^k, s^k)$ in equation (6) can be computed by $\bar{Q}^k(a^k, s^k) = \sum_{m=1}^{M} Q_m N_m^k \pi_m^k$. First, we see that when $k = K$, $J_\mu^K(s^K)$ can be rewritten as:

$$\sum_{m=1}^{M} Q_m N_m^K \pi_{\mu,m}^K - \frac{\lambda}{W_{TCP}^K}\left(\sum_{m=1}^{M} N_m^K \pi_{\mu,m}^K - W_{TCP}^K\right)$$
$$= \sum_{m=1}^{M}\left(Q_m - \frac{\lambda}{W_{TCP}^K}\right) N_m^K \pi_{\mu,m}^K + C(W_{TCP}^K), \quad (16)$$

where $[\pi_{\mu,m}^K, m = 1,...,M] = \mu(s^K)$ denotes the vector of transmission permissions given the policy $\mu(s^K)$. Based on this, the expected utility-to-go

$$J_\mu^K(s^K) = \sum_{m=1}^{M}\left(Q_m - \frac{\lambda}{W_{TCP}^K}\right) N_m^K \pi_{\mu,m}^K + \frac{\lambda M}{W_{TCP}^K}$$ is separable

and also a nondecreasing function of $N_m^K$. Then, by assuming $J_\mu^{k+1}(s^{k+1}) = \sum_{m=1}^{M} J_{\mu,m}^{k+1}(W_{TCP}^{k+1}, N_m^{k+1}) + C^{k+1}(W_{TCP}^{k+1})$ and assume $J_{\mu,m}^{k+1}(W_{TCP}^{k+1}, N_m^{k+1})$ are nondecreasing functions of $N_m^{k+1}$ for all classes, we have

$$J_\mu^k(s^k) = u^k(s^k, \mu(s^k)) + \gamma \sum_{s^{k+1} \in \mathcal{S}} P(s^{k+1} \mid s^k) J_\mu^{k+1}(s^{k+1})$$

$$= \sum_{m=1}^{M}\left(Q_m - \frac{\lambda}{W_{TCP}^k}\right) N_m^k \pi_{\mu,m}^k + \frac{\lambda}{W_{TCP}^k} +$$

$$\gamma \sum_{s^{k+1} \in \mathcal{S}} P(W_{TCP}^{k+1} \mid W_{TCP}^k)\left(\begin{array}{c}\sum_{m=1}^{M} J_{\mu,m}^{k+1}(W_{TCP}^{k+1}, N_m^{k+1}(\pi_{\mu,m}^k)) \\ +C^{k+1}(W_{TCP}^k)\end{array}\right)$$

$$= \sum_{m=1}^{M} J_m^k(W_{TCP}^k, N_m^k) + C^k(W_{TCP}^k), \text{ where}$$

$$J_{\mu,m}^k(W_{TCP}^k, N_m^k) = \left(Q_m - \frac{\lambda}{W_{TCP}^k}\right) N_m^k \pi_{\mu,m}^k +$$
$$\gamma \sum_{s^{k+1} \in \mathcal{S}} P(W_{TCP}^{k+1} \mid W_{TCP}^k) J_{\mu,m}^{k+1}(W_{TCP}^{k+1}, N_m^{k+1}(\pi_{\mu,m}^k))$$
, (17)

and $[\pi_{\mu,m}^k, m = 1,...,M] = \mu(s^k)$. Hence, the expected utility-to-go $J_\mu^k(s^k)$ is also separable and $J_{\mu,m}^k(W_{TCP}^k, N_m^k)$ is also a nondecreasing function of $N_m^k$. By backward induction, $J_\mu^k(s^k), \forall s^k \in \mathcal{S}, k = 1,...,K$ are all separable. ∎

## APPENDIX C

*Proof of Theorem 2:* From equation (9), the optimal policy in time slot $i$ with state $s^i$ is the action that maximize the expected utility. Based on Theorem 1, we can see that when the packets in the buffer are independent, the utility-to-go is separable. From equation (17), the optimization problem becomes
$$[\pi_1^{i*}(s^i),...,\pi_M^{i*}(s^i)] =$$
$$\arg\max_{[\pi_1,...,\pi_M]} \sum_{m=1}^{M}\left(\begin{array}{c}\left(Q_m - \frac{\lambda}{W_{TCP}^i}\right) N_m^i \pi_m + \\ \gamma \sum_{s^{i+1} \in \mathcal{S}} P(W_{TCP}^{i+1} \mid W_{TCP}^i) J_{\mu,m}^{i+1}(W_{TCP}^{i+1}, N_m^{i+1}(\pi_m))\end{array}\right).$$

In other words, we have

$$J_{\mu,m}^i(W_{TCP}^i, N_m^i) =$$
$$\begin{cases} J_1 = \left(Q_m - \frac{\lambda}{W_{TCP}^i}\right) N_m^i + \\ \quad\quad \gamma \sum_{s^{i+1} \in \mathcal{S}} P(W_{TCP}^{i+1} \mid W_{TCP}^i) J_{\mu,m}^{i+1}(W_{TCP}^{i+1}, N_m^{A,i}), \text{ if } \pi_m^{i*} = 1 \\ J_0 = \gamma \sum_{s^{i+1} \in \mathcal{S}} P(W_{TCP}^{i+1} \mid W_{TCP}^i) J_{\mu,m}^{i+1}(W_{TCP}^{i+1}, N_m^i - N_m^{D,i} + N_m^{A,i}) \\ \quad\quad , \text{ if } \pi_m^{i*} = 0 \end{cases}$$
. (18)

Hence, by defining $PM_m^{i*}(s^i) = J_1 - J_0$ in equation (18), the problem becomes an optimal stopping problem [23] that leads to the utility-to-go in equation (9). ∎

## APPENDIX D

*Proof of Theorem 4:* Let $|\mathbf{V}| = 1$, the sufficient condition in equation (15) is satisfied. Without loss of generality, let $\bar{Q}_1 \geq \bar{Q}_2 \geq ... \geq \bar{Q}_h \geq \bar{Q}_{h+1} = \mathcal{Q}$. Suppose that the condition is satisfied when $|\mathbf{V}| \leq h$, i.e.

$$\sum_{n=1}^{h}(\bar{Q}_n)^2 \sum_{n=1}^{h}\Delta\bar{Q}_n \geq \sum_{n=1}^{h}\bar{Q}_n \sum_{n=1}^{h}\bar{Q}_n\Delta\bar{Q}_n. \quad (19)$$

For $|\mathbf{V}| = h+1$, since $\mathcal{Q} \leq \bar{Q}_n, \forall V_n \in \mathbf{V}$, if the sufficient conditions are satisfied, we have priority metrics $\Delta\mathcal{Q} = \sum_{\forall m \in \mathbf{M}_{h+1}^k} Q_m N_{m,h+1} \geq \sum_{\forall m \in \mathbf{M}_n^k} Q_m N_{mn} = \Delta\bar{Q}_n, \forall V_n \in \mathbf{V}$. Hence, we have

$$\sum_{n=1}^{h}\Delta\mathcal{Q}\bar{Q}_n(\bar{Q}_n - \mathcal{Q}) \geq \sum_{n=1}^{h}\mathcal{Q}\Delta\bar{Q}_n(\bar{Q}_n - \mathcal{Q})$$
$$\Rightarrow \mathcal{Q}^2\sum_{n=1}^{h}\Delta\bar{Q}_n + \Delta\mathcal{Q}\sum_{n=1}^{h}\bar{Q}_n^2 \geq \mathcal{Q}\sum_{n=1}^{h}\bar{Q}_n\Delta\bar{Q}_n + \mathcal{Q}\Delta\mathcal{Q}\sum_{n=1}^{h}\bar{Q}_n$$
$$(20)$$

Adding equation (19) and (20), we have,

$$\left(\sum_{n=1}^{h}\bar{Q}_n^2 + \mathcal{Q}^2\right)\left(\sum_{n=1}^{h}\Delta\bar{Q}_n + \Delta\mathcal{Q}\right) \geq$$
$$\left(\sum_{n=1}^{h}\bar{Q}_n + \mathcal{Q}\right)\left(\sum_{n=1}^{h}\bar{Q}_n\Delta\bar{Q}_n + \mathcal{Q}\Delta\mathcal{Q}\right),$$

which satisfy the condition in equation (15). By induction, we prove that the sufficient condition of the lemma is fulfilled for all $|\mathbf{V}| \geq 1$. ∎

TABLE IV. NOMENCLATURE

| | | | |
|---|---|---|---|
| $Q_m$ | Distortion impact of the class $CL_m$. | $\mathbf{N}^k$ | $\mathbf{N}^k = [N_m^k, m = 1, ..., M]$ |
| $N_m^k$ | Number of packets in the transmission buffer of class $CL_m$ in time slot $k$ | $a^k$ | Action at time slot $k$ of the FHMDP problem: $a^k = [\pi_m^k, m = 1, ..., M] \in \{0,1\}^M$ |
| $N_m^{k,A}$ | Number of packets in class $CL_m$ that arrive in time slot $k$ | $s^k$ | State at time slot $k$ of the FHMDP problem: $s^k = \{W_{TCP}^k, \mathbf{N}^k\} \in \mathcal{S}$ |
| $N_m^{k,D}$ | Number of packets in class $CL_m$ whose delay deadline is expired in time slot $k$ | $\lambda$ | Positive Lagrangian multiplier of the FHMDP problem |
| $depth_m^k$ | Maximum distance from class $CL_m$ to the root in the DAG in time slot $k$. | $\gamma$ | Discount factor of the FHMDP problem |
| $\bar{Q}^k$ | Expected multimedia distortion reduction in time slot $k$ | $K$ | Finite horizon of the FHMDP problem |
| $PM_m^k$ | Priority metric of class $CL_m$ in time slot $k$ | $J^k(s^k)$ | Expected utility-to-go in time slot $k$ for the FHMDP problem |
| $\mathbf{x}^k$ | $\mathbf{x}^k = [PM_1^k, ..., PM_M^k] \in \mathbb{R}^M$ | $J_m^k(W_{TCP}^k, N_m^k)$ | Expected utility-to-go component specific to the class $CL_m$ |
| $\pi_m^k$ | Transmission permission for transmitting the packets in class $CL_m$ in time slot $k$ | $\mathbf{V}$ | A set of media-TCP users in the network, $\mathbf{V} = \{V_n, n = 1, ..., V\}$ |
| $W^k$ | Congestion window size in time slot $k$ | $\mathcal{F}^k$ | Jain's fairness index of the media-TCP users in the network |
| $p^k$ | Estimated packet loss rate in time slot $k$ | $\bar{Q}_n^k$ | User $V_n$'s expected multimedia distortion reduction in time slot $k$ |
| $W_{TCP}^k(p^k)$ | Expected TCP window size (network congestion metric) | $\Delta\bar{Q}_n^k$ | Expected distortion reduction variation of user $V_n$ in time slot $k$, $\Delta\bar{Q}_n^k = \left|\bar{Q}_n^{k+1} - \bar{Q}_n^k\right|$ |